# Blue-phase templated fabrication of three-dimensional nanostructures for photonic applications

F. Castles[1]*, F. V. Day[1], S. M. Morris[1], D.-H. Ko[2], D. J. Gardiner[1], M. M. Qasim[1], S. Nosheen[1], P. J. W. Hands[1], S. S. Choi[1], R. H. Friend[2] and H. J. Coles[1]

[1] Centre of Molecular Materials for Photonics and Electronics, Department of Engineering, University of Cambridge, 9 JJ Thomson Avenue, Cambridge CB3 0FA, United Kingdom.

[2] Cavendish Laboratory, University of Cambridge, JJ Thomson Avenue, Cambridge CB3 0HE, United Kingdom.

**A promising approach to the fabrication of materials with nanoscale features is the transfer of liquid-crystalline structure to polymers[1-11]. However, this has not been achieved in systems with full three-dimensional periodicity. Here we demonstrate the fabrication of self-assembled three-dimensional nanostructures by polymer templating blue phase I, a chiral liquid crystal with cubic symmetry. Blue phase I was photopolymerized and the remaining liquid crystal removed to create a porous free-standing cast which retains the chiral three-dimensional structure of the blue phase, yet contains no chiral additive molecules. The cast may in turn be used as a hard template for the fabrication of new materials. By refilling the cast with an achiral nematic liquid crystal, we created templated blue phases which have unprecedented thermal stability in the range -125–125 ºC, and that act both as mirrorless lasers and switchable electro-optic devices. Blue-phase templated materials will facilitate advances in device architectures for photonics applications in particular.**

Materials with nanoscale features are of increasing interest in a wide range of applications[12-14]; their fabrication is a central challenge in the field of nanotechnology. Some materials interact strongly with visible light as a result of nanoscale features that possess a periodicity similar to the wavelength of light in the medium. These are of considerable interest in photonic devices that mould the flow of light[13]. Unfortunately, rather few such materials occur naturally; much recent research has concerned their artificial design and fabrication. However, there exists a class of liquid crystals—blue phases (BPs)—that self-assemble into a three-dimensional (3D) periodic lattice with diamond-like structure, and whose unit cell may be tuned to produce vividly coloured Bragg-like reflections over the entire visible wavelength range[15]. They have been investigated in the context of 3D lasers[16-18], flat-panel displays[19], and as soft templates for the formation of 3D colloidal crystals[20].

Ordered liquid crystal phases appear in materials composed of small elongated organic molecules, which can align locally along a common average direction. In the nematic liquid crystal phase the average direction of alignment is uniform. In chiral liquid crystals—formed from intrinsically chiral molecules or by the addition of a chiral dopant to achiral molecules—further structure may develop. In the chiral nematic, or 'cholesteric', phase, the average direction twists along a single axis in a periodic helicoidal fashion. In the chiral BPs, double twist can lead to a 3D periodic structure; BPI has cubic symmetry[15].

New materials have been created by transferring liquid crystalline order to inorganic solids[12,21] and to polymers[1-11]. Following work by Broer *et al.*, photopolymerization of reactive mesogens is commonly used to stabilize a wide variety of liquid crystalline structures, including the BPs, and to increase functionality[7,22-26]. The reactive mesogens, together with a photoinitiator, can be dissolved in a conventional liquid crystalline material; upon exposure to ultraviolet light they form a polymer network. Guo *et al.*[8] were the first to use the washout re-fill method[5,8-11,27], whereby the unpolymerized material is subsequently washed out by solvent and the polymer network refilled by another liquid crystal, to form hyper-reflective chiral nematic layers. Shopsowitz *et al.* have also shown, in an alternative system, that 1D chiral nematic order may be templated in a free-standing porous material, and have postulated that this may be used as a hard template for the formation of new materials[21]. Here we show, using reactive mesogens, that full 3D order may be templated in a free-standing porous material. Further, we experimentally demonstrate its use as a hard chiral template to create a new material: effectively a BP that contains no chiral additive molecules yet is suitable for narrowband mirrorless lasing and electro-optic devices.

The fabrication process is shown in Fig. 1 (see Methods). Essentially, a cast of the BP was formed using a polymer network. The premixtures from which the nanostructure was formed were composed of 3–5 wt % chiral dopant and 25–50 wt % reactive-mesogen/photoinitiator mixture, dissolved in bimesogenic liquid crystal materials. Bimesogens are known to form unusually stable BPs[28], which were found to withstand the high concentrations of reactive mesogen, and high ultraviolet exposures, required for the process. The mixture was capillary filled between two parallel glass sheets (a cell), forming a film 20 μm thick. At Stage 1, BPI self-assembled upon cooling from the isotropic phase, displaying a characteristic platelet texture[28] when observed in transmission using an optical polarizing microscope (Fig. 1), and a characteristic transmission spectrum[28] (Fig. 2). The cell was then illuminated with ultraviolet light to polymerize the reactive mesogens (Fig. 1, Stage 2). To provide a direct comparison between polymerized and unpolymerized regions, a mask was used such that only the centre of the cell was illuminated. At Stage 3 the cell was placed in acetone, typically for 16–22 h. This removed, by diffusion, the bimesogens, the chiral dopant, and any remaining unpolymerized reactive-mesogen/photoinitiator mixture. The removal could be observed by the reduction in the transmission of light on the polarizing optical microscope, as optically active material is replaced by optically inactive, isotropic acetone, and the weakly birefringent polymer template appeared optically isotropic (Fig 1b, Stages 3&4) due to the symmetry of the cubic lattice. The wash-out was typically continued until a few hours after the whole cell became maximally dark. The polymer structure in the UV-exposed region remained obvious to the naked eye (Fig. 1c, Stage 3). When the cell was removed from acetone, any acetone remaining in the cell was left to evaporate at room temperature. The cell could then be opened and the polymer removed using a razor blade. A solid, freestanding, polymer structure was thus obtained (Fig. 1, Stage 4).

We carried out a number of experiments to confirm conclusively that the polymer retained the 3D structure of the blue phase. An unopened cell was refilled with the material 4-cyano-4'-pentylbiphenyl (5CB), as shown in Fig. 1, Stage 5. Subsequent observations were carried out at room temperature. 5CB naturally forms a room-temperature nematic liquid crystal phase. It is achiral so cannot form a BP by itself. In the non-ultraviolet-exposed regions of the cell, where no polymer remained, a nematic phase was observed, as expected (Fig. 1, Stage 5, and Supplementary Fig. S1). In the ultraviolet-exposed regions, which contained the crosslinked polymer, a BP-like structure was observed. This exhibited an apparently identical platelet texture (Fig. 1b, Stage 1 vs Stage 5), a similar characteristic transmission spectrum (Fig. 2a), and a Kossel diffraction diagram for green platelets that corresponds to the structure of BPI viewed along the [011] crystal direction[29] (Fig. 2b). A number of conclusions are drawn from this. First, the polymer was porous, allowing the liquid crystal to refill the voids. Second, the polymer must retain a 3D structure, since BP-like structure was re-transferred *from* the polymer *to* the achiral nematic liquid crystal at this stage. The wavelength of the selective reflection, and hence the observed colour, suggests the periodicity of the structure is approximately a few hundred nanometres. Third, it is experimentally verified that the polymer may be used as a hard template.

The templated achiral BPs are unusually stable. BPs formed from conventional liquid crystal materials are usually stable over a temperature range of 0.5–2°C. However, we typically find that the temperature range of stability of the templated BP is based on the *nematic* phase range of the refilling material, and is therefore much larger than the intrinsic BPs of conventional materials. To exploit this effect, we refilled the polymer with a large-temperature-range achiral nematic: BL006 (Merck, KGaA), which is known to exhibit a nematic phase from below -20 °C to 113 °C. Fig. 3 demonstrates that the resulting structure is largely unchanged from -125 °C to 125 °C. The maximum transmitted intensity of light was reduced at the lower and higher ends of this range, suggesting a gradual transition to the isotropic phase at high temperatures, and a possible glass transition at low temperatures. Smeared out phase transitions are typical of liquids in confined geometries[30]. Further, the structure was observed to remain unchanged after repeated cycles between -150 °C and 150 °C. It is clear that the system exhibits unprecedented stability when compared to conventional liquid crystal materials (0.5–2 °C range), to bimesogenic mixtures (40 °C range[28]), or to the polymer-stabilised system of Kikuchi *et al*. (60 °C range[26]). The templated BPs remain switchable in an applied electric field, which is of use in birefringence phase modulation photonic devices (see Supplementary Fig. S2).

To confirm further the 3D structure of the template, and to demonstrate the use of such a template in photonic devices, we created a BP-templated laser. Again, an unopened cell was refilled, this time with a mixture of 1 wt % laser dye pyrromethene 597 in 5CB. In order to generate lasing, the BP-templated cell was optically pumped with the second harmonic (532 nm) of an Nd:YAG laser and the intensity of the resulting emission was recorded for a range of excitation energies (see Methods). Above a threshold input energy, the templated BP displayed a clear lasing peak at 565 nm (Fig. 4a), which corresponds to the long-wavelength edge of the template's bandgap (c.f., Fig. 2a), as expected for band-edge lasing. However, no laser emission was observed when the untemplated region was pumped directly (Fig. 4b), where only broadband fluorescence was recorded for equivalent excitation energies. Since both regions contain the laser dye, the observation of single mode, narrow linewidth lasing confirms that the emission is a direct result of the polymer template. The threshold input energy for lasing was found to be 681±2 nJ/pulse, (Fig. 4c). Supporting evidence that the emission from the sample is the result of lasing at a photonic band-edge is shown in Fig. 4d, where it can be seen that the output is right circularly polarized and matches the chirality of the initial structure from which the template was formed.

We have succeeded in refilling the polymer nanostructure with a range of achiral liquid crystals (5CB, 4-cyano-4'-octylbiphenyl, E49, BL006 [all Merck KGaA]), each time forming a BP-like structure. Thus templated BPs, with a wide range of periodicities, and with enhanced stability, are possible where previously they were impossible. One may form a BP device using materials that naturally form a BP, such as the bimesogen mixtures, crosslink the BP structure, wash the resultant, and then refill with materials that have desirable properties for applications, such as high birefringence or high dielectric anisotropy. The coexistence of BP/nematic structures is also made possible (Supplementary Fig. S2).

The templated material is unusual in that it induces chirality, yet the additives, post treatment, are non chiral molecules. This, combined with the demonstrated porosity, suggests it may be enantioselective. Since the periodicity is of the order of the wavelength of light, photonic applications are anticipated, such as the lasing and electro-optic devices demonstrated here. The periodicity is tunable over the visible wavelength range by altering the amount of chiral dopant in the premixture. While we demonstrated the use of the material as a hard template by refilling with an achiral nematic liquid crystal—an organic material—it is expected that the polymer will also template inorganic materials. Refilling the template with a high-refractive index material may lead to interesting optical properties, or refilling with a semiconductor may lead to interesting electronic properties. More generally, we have demonstrated a new method for the self-assembled fabrication of 3D nanostructures which may be of use in diverse applications.

**Methods**

**Premixtures**. Premixtures were composed of chiral dopant BDH1281 (Merck KGaA), reactive-mesogens/photoinitiator mixture UCL-011-K1 (Dainippon Ink & Chemicals Inc.), and bimesogenic liquid crystals of the form FFO*n*OFF (see Supplementary Fig. S3) synthesized in-house. Premixture 1 was composed of 3.9 wt % BDH1281, 28 wt % UCL-011-K1, 17 wt % FFO5OFF, 17 wt % FFO7OFF, 17 wt % FFO9OFF, and 17 wt % FFO11OFF. Premixture 2 was composed of 4.7 wt % BDH1281, 47 wt % UCL-011-K1, 16 wt % FFO7OFF, 16 wt % FFO9OFF, and 16 wt % FFO11OFF. Premixture 1 was refilled with 5CB and Premixture 2 was refilled with BL006 (Merck, KGaA).

**Cells.** Cells were made from two pieces of glass coated with indium tin oxide bonded together using glue which contained 20 μm spacer beads applied at the corners. No surface alignment layer was used.

**Initial BP formation.** BPs were formed by cooling from the isotropic phase using a hot-stage (LTS350, Linkam) and hot-stage controller (TMS94, Linkam). For Fig. 1 and Supplementary Fig. S1, Premixture 1 was cooled from 50 °C to 48 °C at a rate of 0.05 °C/min. To create large platelets, suitable for the analysis of Figs. 2 and 4, Premixture 1 was cooled at 0.1 °C/min to just within the BPI temperature range, held at this temperature for typically 4 h while large platelets were grown, then cooled to 48 °C at 0.1 °C/min. For Fig. 3 and Supplementary Fig. S2, Premixture 2 was cooled from 54 °C to 53 °C at 0.02 °C/min, then from 53 °C to 52.5 °C at 1 °C/min.

**Ultraviolet exposure.** Ultraviolet exposure of Premixture 1 was carried out at 48 °C for 7–8 s using an Omnicure Series 1000 Spot Curing System with 320–500 nm filter (EXFO) with light of intensity 50 W/m$^2$ (measured using a PM 100 Digital Optical Power Meter, Thorlabs). Premixture 2 was exposed at 52.5 °C for 5 s.

**Lasing.** Dye-doped BP-templated samples were pumped using a frequency-doubled Nd:YAG laser (Polaris II, New Wave Research) using 5 ns pulses at a 1 Hz repetition rate. The pump laser output was focused to a ≈ 80 μm diameter spot at the cell. The gain medium was the laser dye pyrromethene 597 (Exciton). Emission from the sample was collected in the forward direction, normal to the cell, using a series of collection optics that delivered the output into a fibre-coupled universal serial bus spectrometer (HR2000, Ocean Optics) with a resolution of 0.3 nm. To determine the polarization of the output, a quarter-wave plate and a polarizer were inserted into the set-up before the spectrometer. All laser measurements were carried out at room temperature. The error on the threshold input energy was calculated using standard procedures by assuming linear fits to the data pre- and post-threshold.

**References**

1 de Gennes, P. G. Possibilites offertes par la reticulation de polymeres en presence d'un cristal liquide. *Phys. Lett. A* **28,** 725-726 (1969).
2 Tsutsui, T. & Tanaka, R. Network polymers with cholesteric liquid-crystalline order prepared from poly(gamma-butyl L-glutamate)-butyl acrylate liquid-crystalline system. *Polymer* **22,** 117-123 (1981).
3 Hasson, C. D., Davis, F. J. & Mitchell, G. R. Imprinting chiral structures on liquid crystalline elastomers. *Chem. Commun.* **22,** 2515-2516 (1998).
4 Mao, Y. & Warner, M. Imprinted networks as chiral pumps. *Phys. Rev. Lett.* **86,** 5309-5312 (2001).
5 Jakli, A., Nair, G. G., Lee, C. K., Sun, R. & Chien, L. C. Macroscopic chirality of a liquid crystal from nonchiral molecules. *Phys. Rev. E* **63,** 061710 (2001).
6 Courty, S., Tajbakhsh, A. R. & Terentjev, E. M. Stereo-selective swelling of imprinted cholesteric networks. *Phys. Rev. Lett.* **91,** 085503 (2003).
7 Mitov, M. & Dessaud, N. Going beyond the reflectance limit of cholesteric liquid crystals. *Nature Mater.* **5,** 361-364 (2006).
8 Guo, J. *et al.* Polymer stabilized liquid crystal films reflecting both right- and left-circularly polarized light. *Appl. Phys. Lett.* **93,** 201901 (2008).
9 McConney, M. E. *et al.* Thermally induced, multicolored hyper-reflective cholesteric liquid crystals. *Adv. Mater.* **23,** 1453-1457 (2011).
10 McConney, M. E., Tondiglia, V. P., Hurtubise, J. M., White, T. J. & Bunning, T. J. Photoinduced hyper-reflective cholesteric liquid crystals enabled via surface initiated photopolymerization. *Chem. Commun.* **47,** 505-507 (2011).
11 McConney, M. E. *et al.* Dynamic high contrast reflective coloration from responsive polymer/cholesteric liquid crystal architectures. *Soft Matter* **8,** 318-323 (2012).
12 Kresge, C. T., Leonowicz, M. E., Roth, W. J., Vartuli, J. C. & Beck, J. S. Ordered mesoporous molecular sieves synthesized by a liquid-crystal template mechanism. *Nature* **359,** 710-712 (1992).
13 Joannopoulos, J. D., Villeneuve, P. R. & Fan, S. H. Photonic crystals: putting a new twist on light. *Nature* **386,** 143-149 (1997).
14 Soukoulis, C. M. & Wegener, M. Past achievements and future challenges in the development of three-dimensional photonic metamaterials. *Nature Photon.* **5,** 523-530 (2011).
15 Wright, D. C. & Mermin, N. D. Crystalline liquids: the blue phases. *Rev. Mod. Phys.* **61,** 385-432 (1989).
16 Cao, W., Munoz, A., Palffy-Muhoray, P. & Taheri, B. Lasing in a three-dimensional photonic crystal of the liquid crystal blue phase II. *Nature Mater.* **1,** 111-113 (2002).
17 Yokoyama, S., Mashiko, S., Kikuchi, H., Uchida, K. & Nagamura, T. Laser emission from a polymer-stabilized liquid-crystalline blue phase. *Adv. Mater.* **18,** 48-51 (2006).
18 Coles, H. & Morris, S. Liquid-crystal lasers. *Nature Photon.* **4,** 676-685 (2010).
19 Hisakado, Y., Kikuchi, H., Nagamura, T. & Kajiyama, T. Large electro-optic Kerr effect in polymer-stabilized liquid-crystalline blue phases. *Adv. Mater.* **17,** 96-98 (2005).
20 Ravnik, M., Alexander, G. P., Yeomans, J. M. & Zumer, S. Three-dimensional colloidal crystals in liquid crystalline blue phases. *P. Natl. Acad. Sci. U.S.A.* **108,** 5188-5192 (2011).
21 Shopsowitz, K. E., Qi, H., Hamad, W. Y. & MacLachlan, M. J. Free-standing mesoporous silica films with tunable chiral nematic structures. *Nature* **468,** 422-425 (2010).
22 Broer, D. J., Finkelmann, H. & Kondo, K. In-situ photopolymerization of an oriented liquid-crystalline acrylate. *Makromol. Chem.* **189,** 185-194 (1988).
23 Kitzerow, H.-S. *et al.* Observation of blue phases in chiral networks. *Liq. Cryst.* **14,** 911-916 (1993).

24      Broer, D. J., Lub, J. & Mol, G. N. Wide-band reflective polarizers from cholesteric polymer networks with a pitch gradient. *Nature* **378,** 467-469 (1995).
25      Dierking, I. Polymer network-stabilized liquid crystals. *Adv. Mater.* **12,** 167-181 (2000).
26      Kikuchi, H., Yokota, M., Hisakado, Y., Yang, H. & Kajiyama, T. Polymer-stabilized liquid crystal blue phases. *Nature Mater.* **1,** 64-68 (2002).
27      Choi, S. S., Morris, S. M., Huck, W. T. S. & Coles, H. J. Simultaneous red-green-blue reflection and wavelength tuning from an achiral liquid crystal and a polymer template. *Adv. Mater.* **22,** 53-56 (2010).
28      Coles, H. J. & Pivnenko, M. N. Liquid crystal 'blue phases' with a wide temperature range. *Nature* **436,** 997-1000 (2005).
29      Miller, R. J. & Gleeson, H. F. Lattice parameter measurements from the Kossel diagrams of the cubic liquid crystal blue phases. *J. Phys. II France* **6,** 909-922 (1996).
30      Hikmet, R. A. M. in *Liquid Crystals in Complex Geometries* Ch. 3  (eds G. P. Crawford & S. Zumer) 53-82 (Taylor & Francis, London, 1996).

**Acknowledgements**
This work was carried out under the COSMOS project which is funded by the Engineering and Physical Sciences Research Council UK (Grants No. EP/D04894X/1 and No. EP/H046658/1). We thank H. Hasebe (Dainippon Ink & Chemicals Inc., Japan) for supplying the reactive mesogen UCL-11-K1. S.S.C. acknowledges LG Display for a studentship. S.M.M. acknowledges The Royal Society for financial support.

**Author Contributions**
F.C. conceived the idea. F.C., F.V.D., and S.S.C. developed the fabrication process. F.C. and F.V.D. carried out the microscopy and spectroscopy experiments. F.C. and S.M.M. carried out the Kossel diffraction experiment. S.M.M. and F.C. carried out the lasing experiment. F.C., D.H.K., and D.J.G. characterized the templated structures. M.M.Q. and S.N. synthesized the bimesogenic materials. F.C., P.J.W.H., and F.V.D. fabricated the glass cells. F.C. and F.V.D. wrote the paper in collaboration with all the authors. R.H.F. was a collaborator on the COSMOS project. H.J.C. informed and directed the research.

**Additional information**
The authors declare no competing financial interests. Supplementary information accompanies this paper on www.nature.com/naturematerials. Reprints and permissions information is available online at http://www.nature.com/reprints. Correspondence and requests for materials should be addressed to H.J.C.

**Figure legends**

Figure 1. **Formation of the 3D nanostructured polymer**. **a**, Schematic diagram of the procedure. **b**, Transmission optical polarizing microscopy images (100 μm scale bar). **c**, Photographs of cell (5 mm scale bars). Stage 1: a blue phase self-assembles between two glass sheets. Stage 2: the cell is exposed to ultraviolet light with a square-aperture mask to photopolymerize the reactive mesogens. Upon cooling to room temperature, the exposed region is observed to be stabilised in the blue phase, while the unexposed regions transition to the chiral nematic phase. Stage 3: the cell is placed in acetone to wash out the liquid crystal, chiral dopant, and the remaining reactive mesogen/photoinitiator mixture. The remaining polymer structure exhibits low transmission between crossed polarizers. Stage 4: the solid polymer structure may be removed from the cell. Stage 5: an unopened cell is refilled with the achiral nematic liquid crystal 5CB. There are no chiral molecules included as additives, yet a blue-phase-like structure is observed in the polymer templated regions.

Figure 2. **Optical characterization.** **a**, The transmission spectrum observed before photopolymerization (Stage 1) is characteristic of the blue phase[28]. This shape is preserved upon photopolymerization (Stage 2). After the non-polymerised material is removed, no selective reflection is observed (Stage 3). Upon refilling with a nematic material, the characteristic shape is recovered (Stage 5). **b**, A Kossel diagram at Stage 5 using light of wavelength 405 nm.

Figure 3. **Thermal stability of the templated blue phase.** Polarizing optical microscope images indicate the range of thermal stability (100 μm scale bar). **a**, Images with equal illumination indicate gradual loss of birefringence at low and high temperatures. **b**, Intensity enhanced images clarify that the structure persists over a range of at least 250 ºC, from -125 °C to 125 °C.

Figure 4. **Band-edge lasing from a templated blue phase. a,** Laser emission from the dye-doped BP-templated region. **b**, Fluorescence from the untemplated region. **c**, Output intensity as a function of the excitation energy for the laser emission in the BP-templated region. Error bars were obtained from the experimental standard deviation of 12 repeated measurements. **d**, Emission spectra for right and left circularly polarized light in the BP-templated region. (200 μm scale bars.)

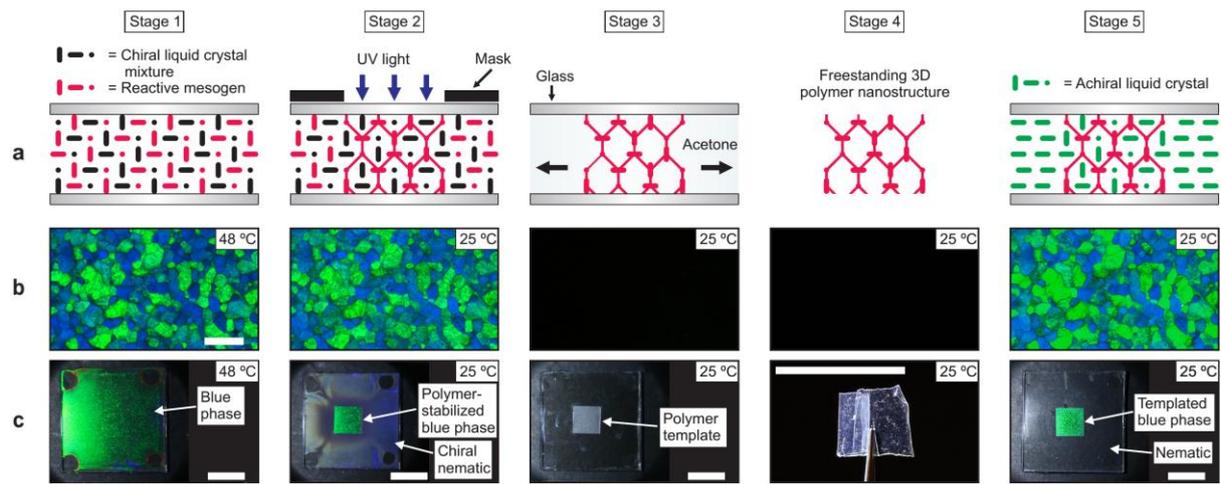

Figure 1.

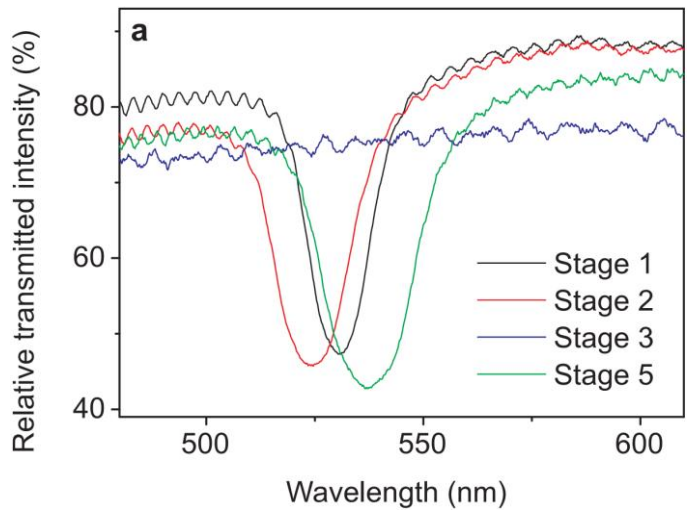 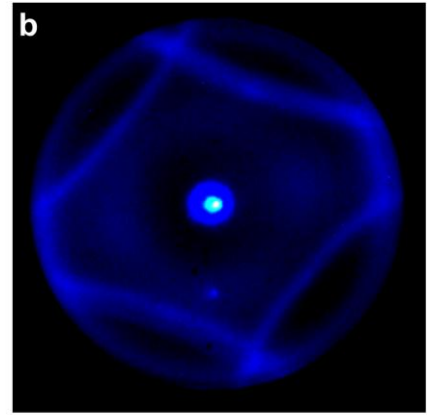

Figure 2.

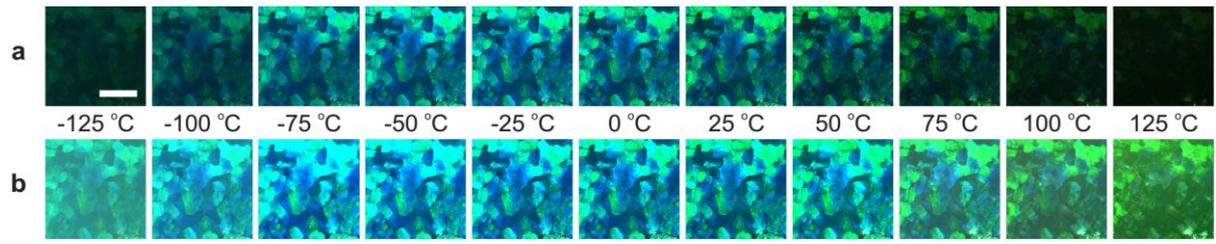

Figure 3.

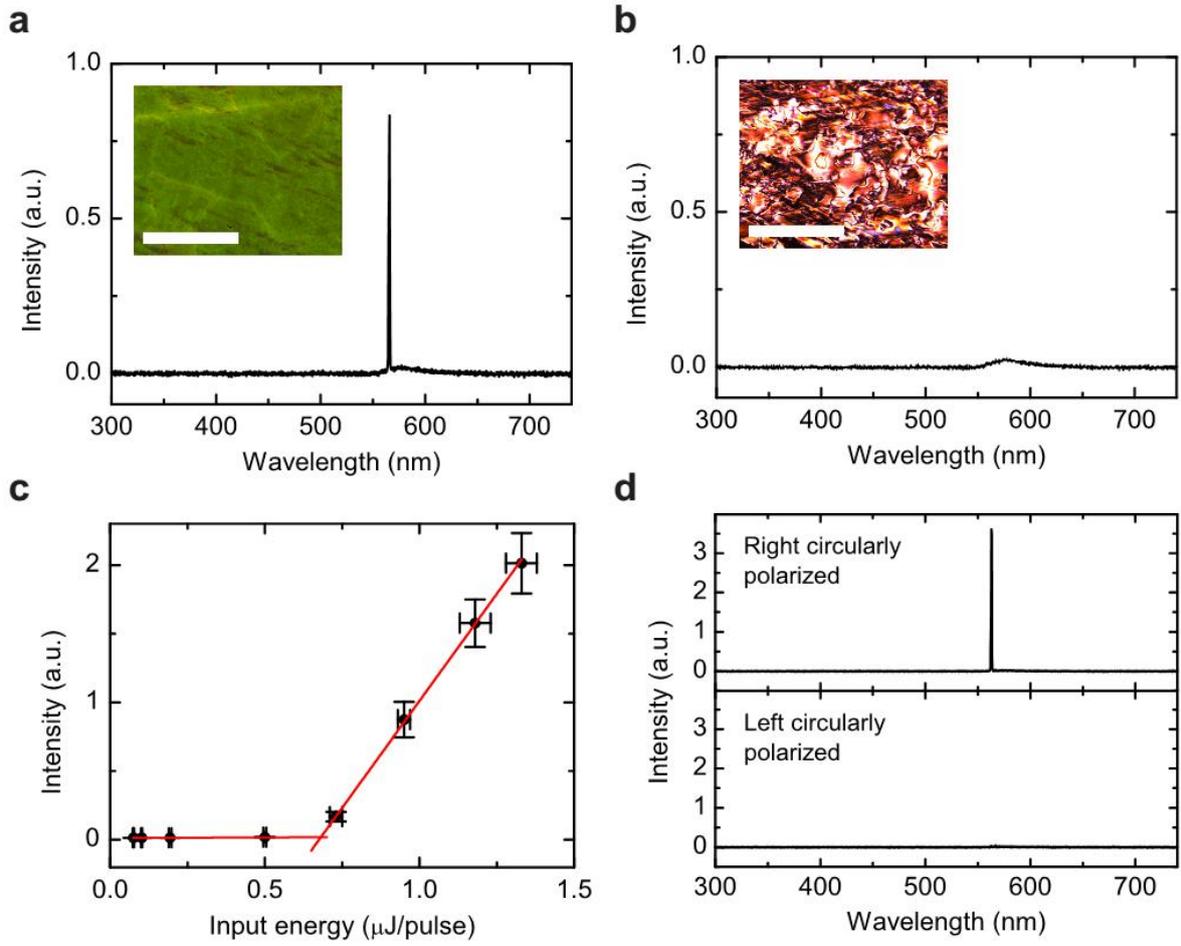

Figure 4.